# Heavily Fe-doped *n*-type ferromagnetic semiconductor (In,Fe)Sb with high Curie temperature and large magnetic anisotropy

Nguyen Thanh Tu,[1,2,] Pham Nam Hai,[3,4,] Le Duc Anh,[1,5] and Masaaki Tanaka[1,4,]

[1]*Department of Electrical Engineering & Information Systems, The University of Tokyo,*
*7-3-1 Hongo, Bunkyo, Tokyo 113-8656, Japan.*
[2]*Department of Physics, Ho Chi Minh City University of Pedagogy,*
*280, An Duong Vuong Street, District 5, Ho Chi Minh City 748242, Vietnam*
[3]*Department of Electrical and Electronic Engineering, Tokyo Institute of Technology,*
*2-12-1 Ookayama, Meguro, Tokyo 152-0033, Japan.*
[4]*Center for Spintronics Research Network (CSRN), The University of Tokyo,*
*7-3-1 Hongo, Bunkyo, Tokyo 113-8656, Japan.*
[5] *Institute of Engineering Innovation, The University of Tokyo,*
*7-3-1 Hongo, Bunkyo, Tokyo 113-8656, Japan.*

We present high-temperature ferromagnetism and large magnetic anisotropy in heavily Fe-doped *n*-type ferromagnetic semiconductor $(In_{1-x},Fe_x)Sb$ ($x = 20 - 35\%$) thin films grown by low-temperature molecular beam epitaxy. The $(In_{1-x},Fe_x)Sb$ thin films with $x = 20 - 35\%$ maintain the zinc-blende crystal and band structure with single-phase ferromagnetism. The Curie temperature ($T_C$) of $(In_{1-x},Fe_x)Sb$ reaches 390 K at $x = 35\%$, which is significantly higher than room temperature and the highest value so far reported in III-V based ferromagnetic semiconductors. Moreover, large coercive force ($H_C = 160$ Oe) and large remanent magnetization ($M_r/M_S = 71\%$) have been observed for a $(In_{1-x},Fe_x)Sb$ thin film with $x = 35\%$. Our results indicate that the *n*-type ferromagnetic semiconductor $(In_{1-x},Fe_x)Sb$ is very promising for spintronics devices operating at room temperature.

Carrier-induced ferromagnetic semiconductors (FMSs) are unique materials with both ferromagnetic and semiconducting properties. By using FMSs we can design new devices with spin-dependent output characteristics such as spin diodes,[1] and spin transistors,[2] which are very promising for non-volatile and low-power-consumption spintronics devices. To be used for realistic devices, however, FMSs must satisfy several fundamental requirements. First, their Curie temperature ($T_C$) must be much higher than room temperature (300 K). Second, both *p*-type and *n*-type FMSs are needed to make *p-n* junctions. Third, the magnetic anisotropy of FMSs must be large enough. Over the past years, however, many efforts on various FMS materials have failed to realize a FMS that can satisfy these three requirements. The prototypical Mn-based FMSs, such as (Ga,Mn)As and (In,Mn)As, have relatively large magnetic anisotropy,[3-6] however, their highest $T_C$ value (200 K) is still much lower than room temperature (300 K)[7,8] and there is no *n*-type Mn-doped III-V FMS available, making it difficult to realize ferromagnetic *p-n* junctions. Recently, we have realized a new class of FMSs using narrow-gap semiconductors, such as InAs, GaSb, and InSb, as the host semiconductors, and Fe as the magnetic dopant. Since Fe atoms replacing group III cation sites are in the neutral state ($Fe^{3+}$) in III-V semiconductors, we are able grow both *p*-type (Ga,Fe)Sb[9-11], insulating (lightly *p*-type) (Al,Fe)Sb[12], *n*-type (In,Fe)As[13-15] and (In,Fe)Sb.[16,17] We have recently obtained $T_C$ = 340 K in $(Ga_{1-x},Fe_x)Sb$ with $x$ = 25%[11] and $T_C$ = 335 K in $(In_{1-x},Fe_x)Sb$ with $x$ = 16%,[16] which are the highest $T_C$ values reported in III-V FMSs so far. However, Fe-doped FMSs lack large magnetic anisotropy, thus they have small remanent magnetization and small coercive force, which are inappropriate for device applications. Furthermore, the highest $T_C$ (335 - 340 K) obtained so far is not high enough for stable operation of spin devices at room temperature (300 K). Therefore, improvement on both the magnetic anisotropy and $T_C$ are strongly required. In this letter, to enhance the magnetic anisotropy and $T_C$ of the $n$-type FMS (In,Fe)Sb, we studied $(In_{1-x},Fe_x)Sb$ thin films with high Fe concentrations ($x$ = 20 - 35%). If $T_C$ depends on $x$ and the hole (or electron) concentration $p$ (or $n$) as $T_C \propto xp^{1/3}$ (or $xn^{1/3}$) [according to the mean-field theory in Refs. 18 and 19], increasing $x$ is the most straightforward way to increase $T_C$. We found that $T_C$ of $(In_{1-x},Fe_x)Sb$ films reaches up to 390 K at $x$ = 35% and the magnetic anisotropy of (In,Fe)Sb can be significantly enhanced by increasing $x$.

Figure 1(a) and Table 1 show the schematic structure and parameters of our samples, respectively. The studied $(In_{1-x},Fe_x)Sb$ layers were grown on semi-insulating GaAs(001) substrates by low-temperature molecular beam epitaxy (LT-MBE). To prevent precipitation of Fe atoms, the $(In_{1-x},Fe_x)Sb$ layers with $x$ = 20 – 35% were grown at a low substrate temperature $T_S$ = 200 – 220°C and its thickness was reduced from 30 nm (at $x$ = 20%) to 12 nm (at $x$ = 30 and 35%) as shown in the 3rd and 4th columns of Table I. Here, the Fe flux was calibrated by secondary ion mass spectroscopy (SIMS) and Rutherford back scattering (RBS). The *in situ* reflection high-energy electron diffraction (RHEED) patterns observed during the MBE growth and X-ray diffraction (XRD) spectra of the (In,Fe)Sb layers show zinc-blende-type crystal structure without any visible second phase. This means that our (In,Fe)Sb thin films maintain the zinc-blende crystal structure of the host semiconductor and we confirmed successful growth of (In,Fe)Sb alloys [see Supplementary Information Section 1 for details of the MBE growth, RHEED patterns and XRD spectra].

We investigate the magneto-optical properties of our $(In_{1-x},Fe_x)Sb$ samples by magnetic circular dichroism (MCD) spectroscopy in a reflection setup [see

Supplementary Information Section 2 for details of the MCD measurements]. Figures 1(c) – (e) show the normalized MCD spectra of sample B1, B2 and B3 (($In_{1-x},Fe_x$)Sb with $x$ = 20, 30 and 35%), respectively, measured at 5 K with various magnetic fields ($H$ = 0.2, 0.5, and 1 T) applied perpendicular to the film plane. Here, the intensity of the MCD spectra measured at 0.2 and 0.5 T were normalized by the intensity at $E_1$ = 1.75 eV of the spectrum measured at 1 T. For a reference, we also show the MCD spectra of an undoped InSb and ($In_{1-x},Fe_x$)Sb samples with $x$ = 5% and 11% in Fig. 1(b).[16] In all MCD spectra of sample B1 – B3, there is no broad background, which would be observed if metallic Fe or Fe-related precipitates exist. Thus, we can rule out the existence of Fe-related second-phase precipitates in our ($In_{1-x},Fe_x$)Sb samples. The normalized MCD spectra measured with various magnetic fields show almost overlapping on a single spectrum in the whole photon-energy range, indicating that all the MCD peaks come from single-phase ferromagnetism in (In,Fe)Sb. The spectral shape of sample B1 – B3 are quite similar to those of the undoped InSb and ($In_{1-x},Fe_x$)Sb samples with $x$ = 5% and 11%. We observed MCD peaks at $E_1$ = 1.75 eV and $E_1 + \Delta_1$ = 2.7 eV, which originate from the optical critical point energy peaks $E_1$ (2 eV) and $E_1 + \Delta_1$ (2.4 eV) of the InSb band structure, respectively.[20] The reason for the red shift (~0.25 eV) of $E_1$ and the blue shift (~0.3 eV) of $E_1 + \Delta_1$ are not clear at present. They may be due to the modification of the band gap energy of InSb by heavy Fe doping ($x \geq 20\%$), and/or the optical interference effect as observed in the MCD spectra of thin FMS films such as GaMnAs film (10–100 nm)[21] or (In,Fe)As film (10 nm).[22] Figures 1(f) – 1(h) show the MCD intensity – magnetic field (MCD – $H$) curves of sample B1 – B3 ($x$ = 20 – 35%), measured at photon energy $E_1$ = 1.75 eV at various temperatures $T$ = 5 – 320 K. Here, we cannot obtain the data at temperature $T > 320$ K due to the limitation of our MCD measurement system. All the samples show open hysteresis curves even at 320 K, indicating that $T_C$ is higher than 320 K. As $x$ increases, the MCD intensity saturates at a higher magnetic field, suggesting that the in-plane magnetic anisotropy become more dominant in these samples.

Figures 2(a) – 2(c) show the magnetization *vs.* magnetic field ($M - H$) curves of sample B1 – B3 ($x$ = 20 – 35%), measured at 10 K by superconducting quantum interference device (SQUID) magnetometry with a magnetic field applied in the film plane along the [110] axis and perpendicular to the film plane along the [001] axis, respectively. For the sample with $x$ = 20%, the in-plane magnetization and perpendicular magnetization show similar characteristics, suggesting small magnetic anisotropy in this sample. For the samples with $x$ = 30 and 35%, however, the difference between in-plane and perpendicular magnetizations is remarkable. The in-plane magnetization saturates at a much smaller magnetic field than the perpendicular magnetization, and the in-plane remanent magnetization ($M_r$) is much larger than the perpendicular $M_r$. This result suggests that the easy axis of magnetization is in the film plane and there is a strong magnetic anisotropy in these samples. In Figs. 2(a) – 2(c), the saturation magnetization ($M_S$) values are 110, 145 and 165 emu/cc (corresponding to 3.6, 3.4 and 3.2 $\mu_B$/Fe atom) at $x$ = 20, 30, and 35%, respectively; and the saturation field ($H_S$), determined by the intersection of the in-plane $M$-$H$ loop and the perpendicular $M$-$H$ loop (black arrows), are 0.3 T, 0.5 T, and 0.5 T for the samples with $x$ = 20, 30, and 35%, respectively. From these values, we estimated the anisotropy constant $K_u = \int_0^{H_S} (M_{//} - M_{\perp}) dH - 2\pi M_S^2$, where $M_{//}$ and $M_{\perp}$ stand for in-plane and perpendicular magnetization, respectively. We found that $K_u$ is negligible at $x$ = 20% but increases to $3.9 \times 10^4$ erg/cc at $x$ = 30% and $5.3 \times 10^4$ erg/cc at $x$ = 35%, indicating that the magnetic

anisotropy can be enhanced by increasing $x$. Figure 2(d) shows the enlarged in-plane $M$-$H$ curves of sample B1 – B3 ($x$ = 20 - 35%) near zero field at 10 K. One can see that the coercive force ($H_C$) increases from 80 Oe (at $x$ = 20%) to 160 Oe (at $x$ = 30% and 35%). Furthermore, the remanent magnetization ($M_r$) increases from 15 emu/cc ($M_r/M_S$ = 16%) at $x$ = 20% to 117 emu/cc ($M_r/M_S$ = 71%) at $x$ = 35%. This result indicates that $H_C$ and $M_r$ are drastically increased by increasing $x$. It should be recalled that the big problem of Fe-doped FMSs so far is the small $H_C$ and $M_r$. Therefore, the large $H_C$ (160 Oe) and large $M_r$ (65%) observed in (In,Fe)Sb samples $x$ = 35% is an important step towards practical spintronics devices.

In Fig. 3(a), we show the field cooling (FC, closed circles) and zero field cooling (ZFC, open circles) $M - T$ curves of sample B1 – B3 ($x$ = 20 – 35%). Before the $M - T$ measurement, samples were cooled from room temperature to 10 K with an in-plane magnetic field of 1 T for the FC process and without magnetic field for the ZFC process. To obtain clear signals, a small in-plane magnetic field of 30 Oe was applied during the magnetization ($M - T$) measurement when increasing temperature from 10 K to 400 K. The $M - T$ curves of the samples with $x$ = 20 – 35% show clear increase of the magnetization as $x$ increases, consistent with the $M - H$ curves. In all the $M - T$ curves shown in Fig. 3 (a), there is a separation between FC and ZFC at low temperature, indicating the blocking temperature $T_B$ = 15, 55, and 85 K for the samples with $x$ = 20, 25, and 35%, respectively. This suggests that there are superparamagnetic Fe-rich (In,Fe)Sb domains in our samples. Considering the high Fe doping concentration $x$ in an InSb matrix, spinodal decomposition[23-26] can occur and cause a local fluctuation of Fe concentration while maintaining the zinc-blende crystal structure of the host semiconductor InSb, as in the case of many other FMSs.[27, 28, 10] Note that such local fluctuation of Fe concentration has been reported in the (In,Fe)Sb sample $x = 16\%$,[16] thus it is not surprising to observe this behavior in the samples with $x = 20 - 35\%$. Figure 3 (b) shows the magnetic states in our samples at a temperature lower than $T_B$ in the FC and ZFC processes. Here, the small blue arrows and big red arrows show magnetic moments of Fe atoms and the total magnetic moments of Fe-rich (In,Fe)Sb domains (blue areas), respectively. The superparamagnetic behavior appears due to the existence of unconnected Fe-rich (In,Fe)Sb domains.

We then estimated $T_C$ of the (In,Fe)Sb samples by using the Curie-Weiss plot at high temperatures, because the disappearance of $M$ is not clear in the $M - T$ curves (around 350 – 400 K) due to the small applied magnetic field (30 Oe) (see Supplementary Information Section 3 for details of the Curie-Weiss plot). $T_C$ values are listed in the 6th column of Table I and plotted as function of $x$ (red diamonds) in Fig. 3(c). For comparison, $T_C$ of the $(In_{1-x},Fe_x)Sb$ samples with $x = 1 – 16\%$ are also shown (black circles).[16] One can see that $T_C$ increased from 335 K at $x = 16\%$[16] to 390 K at $x = 35\%$, which is far above room temperature (300 K), and thus promising for realistic devices operating at room temperature. The reason why $T_C$ slowly increases in the range of $x = 20 – 35\%$ is unclear at the present. However, we found that this was accompanied by a sudden change in the lattice constant – doped Fe concentration relation, which was confirmed by our XRD measurement [see Supplementary Information Section 1 for details of lattice constant]. As $x$ increase, the lattice constant $a$ of the heavily Fe-doped samples ($x = 20 – 35\%$) does not decrease like that of samples with $x = 1 – 16\%$. Therefore, we suggest that a part of Fe atoms may reside at the interstitial sites in heavily Fe-doped samples, and such interstitial Fe atoms prefer anti-ferromagnetic coupling. which leads to the slowdown of increasing $T_C$ due to competition between anti-ferromagnetic coupling and ferromagnetic coupling.

In Fig. 3(d), we show the temperature dependence of the electrical resistivity $\rho$ of the sample B1 – B3 ($x$ = 20 – 35%). The resistivity $\rho$ was measured in the temperature range from 5 to 300 K using patterned Hall bars (length 200 μm, width 50 μm). The $\rho$ values measured at 300 K are listed in the 7th columns of Table I. One can see that $\rho$ decreases as $x$ increases. Sample B1 ($x$ = 20%) shows insulating behavior, while sample B2 ($x$ = 30%) and B3 ($x$ = 35%) show nearly metallic behavior: At low temperature, $\rho$ of sample B2 and B3 is three to four orders of magnitude smaller than that of sample B1. Interestingly, the insulator-metal-transition near $x \sim$ 30% coincides with the sudden increase of $K_u$, $H_C$, and $M_r$ of sample B2 and B3 ($x$ = 30% and 35%), suggesting a drastic change in the electronic band structure of $(In_{1-x},Fe_x)Sb$ at $x \geq$ 30%. Based on these facts, we suggest the following scenario for the transition of the electronic band structure in (In,Fe)Sb. First, we recall that all the (In,Fe)Sb samples with $x \leq$ 20% are insulating at low temperatures with low electron concentrations $n \sim 10^{17}$ cm$^{-3}$, indicating that the Fe atoms are in the isoelectronic $Fe^{3+}$ states. Thus, the Fe $d$ electrons are localized with the total orbital momentum $L \sim$ 0 for their orbital $d^5$ configuration, meaning that their magnetic anisotropy energy $\lambda S \cdot L \sim$ 0 (here, $S$ is the total spin quantum number of the Fe $d$ electrons and $\lambda$ is the strength of the spin-orbit interaction). Meanwhile, the free electrons in the conduction band have wave functions mainly composed of the $s$ orbitals of the In atoms with the orbital momentum $l \sim$ 0. Thus, the free electrons do not contribute to the magnetic anisotropy. This explains the very small magnetic anisotropy in $n$-type Fe-doped narrow-gap FMSs, such as (In,Fe)As and (In,Fe)Sb with the $Fe^{3+}$ states. For the (In,Fe)Sb samples with $x \geq$ 30%, however, the metallic conduction suggests that the Fe $d$ electrons may become itinerant, thus $L$ of Fe $d$ electrons and their magnetic anisotropy energy $\lambda S \cdot L$ may be non-zero, explaining the sudden increase of $K_u$, $H_C$ and $M_r$ of these samples. Further investigations of their electronic structures by photoemission spectroscopy will be needed to confirm our explanation.

In summary, we have grown (In,Fe)Sb thin films with high Fe concentrations $x$ = 20, 30 and 35%. Magnetic properties investigated by MCD and SQUID measurements demonstrate single-phase ferromagnetism with high $T_C$ and large magnetic anisotropy. The (In$_{1-x}$,Fe$_x$)Sb thin film with $x$ = 35% shows $T_C$ as high as 390 K, which is far above room temperature, large coercive force ($H_C$ = 160 Oe), and large remanent magnetization ($M_r/M_S$ = 71%). Our results indicate that $n$-type FMS (In,Fe)Sb is very promising for semiconductor spintronics devices operating at room temperature.

**Acknowledgments**
This work is supported by Grants-in-Aid for Scientific Research (Grant No. 23000010, 16H02095, 15H03988, 17H04922), CREST of JST (No. JPMJCR1777), the Yazaki Foundation, and the Spintronics Research Network of Japan. N. T. T. acknowledges support from the JSPS Postdoctoral Fellowship Program (No. P15362).

**Table I.** Thickness $d$, growth temperature $T_{sub}$, anisotropy constant $K_u$, Curie temperature $T_C$, and resistivity $\rho$, of $(In_{1-x},Fe_x)Sb$ thin films of samples B1, B2 and B3 with various Fe concentrations $x$ = 20, 30, and 35%

| Sample | $x$ % | $d$ (nm) | $T_{sub}$ (C) | $K_u$ (erg/cc) | $T_C$ (K) | $\rho$ (Ωcm) |
|---|---|---|---|---|---|---|
| B1 | 20 | 30 | 220 | - | 360 ± 5 K | $3.2 \times 10^{-2}$ |
| B2 | 30 | 12 | 200 | $3.9 \times 10^4$ | 380 ± 5 K | $7.5 \times 10^{-3}$ |
| B3 | 35 | 12 | 200 | $5.3 \times 10^4$ | 390 ± 5 K | $6.6 \times 10^{-3}$ |

**Figure captions**

**Figure 1.** (a) Schematic sample structure studied in this work. (b) Reflection MCD spectra measured at 5 K under a magnetic field of 1 T applied perpendicular to the film plane for an undoped InSb and $(In_{1-x},Fe_x)Sb$ samples with $x$ = 5 and 11%. (c) – (e) Normalized reflection MCD spectra measured at 5 K under various magnetic fields of $H$ = 0.2, 0.5, and 1 T applied perpendicular to the film plane for $(In_{1-x},Fe_x)Sb$ samples B1 – B3 ($x$ = 20, 30, 35%), respectively. Here, the intensity of the MCD spectra measured at 0.2 and 0.5 T are normalized by the intensity at $E_1$ = 1.75 eV of the spectrum measured at 1 T. (f) – (h) MCD – $H$ characteristics measured at 5 – 320 K at a photon energy of $E_1$ = 1.75 eV of samples B1 – B3 ($x$ = 20 – 35%), respectively.

**Figure 2.** (a) - (c) Magnetization *vs.* magnetic field ($M – H$) curves of samples B1 – B3 ($x$ = 20 - 35%), measured at 10 K with a magnetic field applied in the film plane along the [110] axis and perpendicular to the film plane along the [001] axis, respectively. (d) $M – H$ curves of the same samples near zero field at 10 K when the magnetic field was applied in the film plane along the [110] axis. The arrows in (a) – (c) indicate the position of the saturation magnetic field $H_S$.

**Figure 3.** (a) Temperature dependence of the magnetization ($M – T$ curves) of samples B1 – B3 ($x$ = 20 – 35%). The samples were cooled from room temperature to 10 K under two conditions, with a magnetic field of 1 T (FC, closed circles) and zero magnetic field (ZFC, open circles). After cooling, the magnetization was measured with increasing temperature under a weak magnetic field of 30 Oe applied in the film plane along the [110] direction. (b) Schematic pictures of the magnetic states in the (In,Fe)Sb

samples at $T < T_B$ in FC and ZFC processes, respectively. (c) $T_C$ of $(In_{1-x},Fe_x)Sb$ as a function of $x$. (d) Temperature dependence of the electrical resistivity $\rho$ for samples B1 – B3 ($x = 20 - 35\%$).

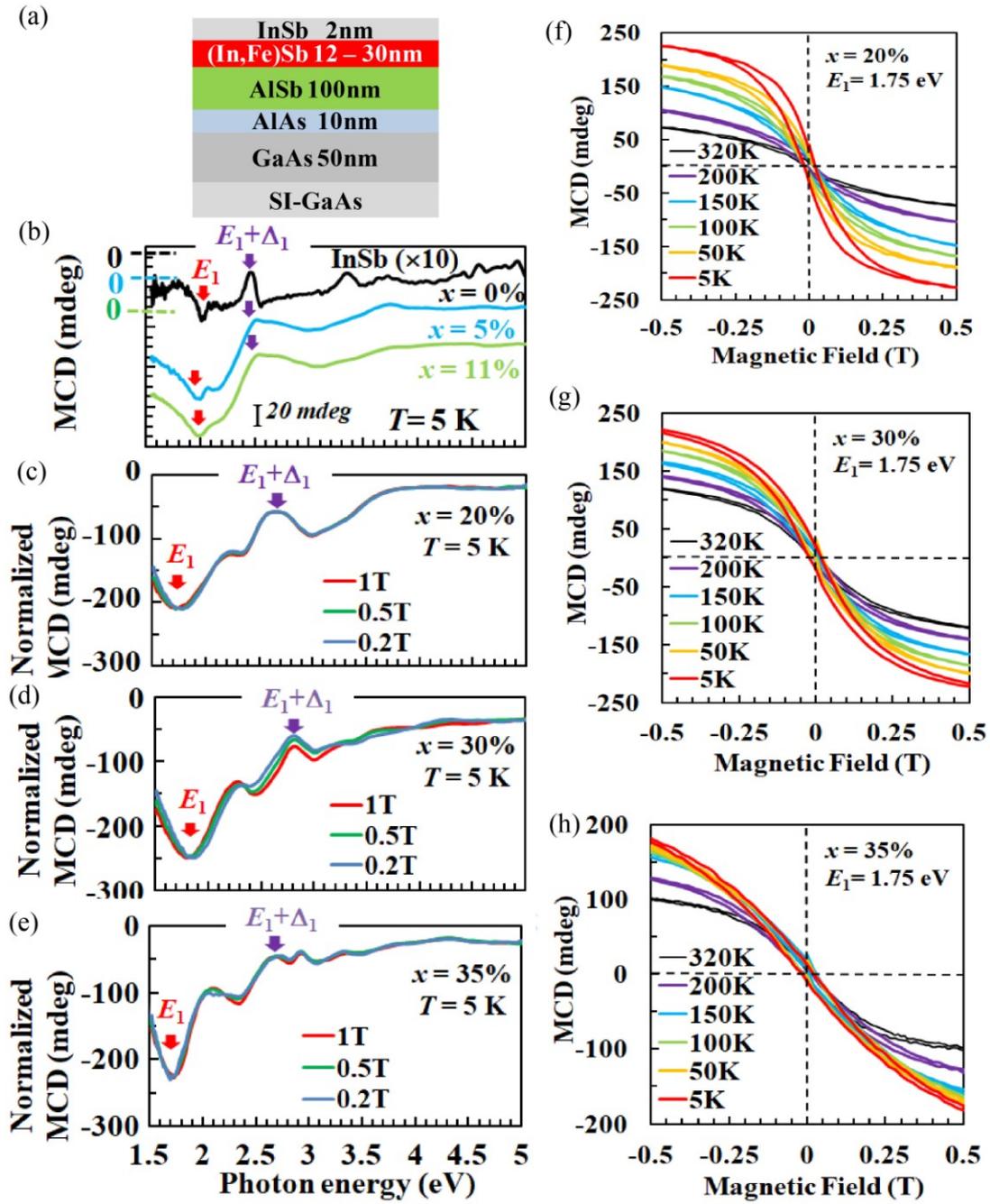

Fig. 1. Tu *et al*

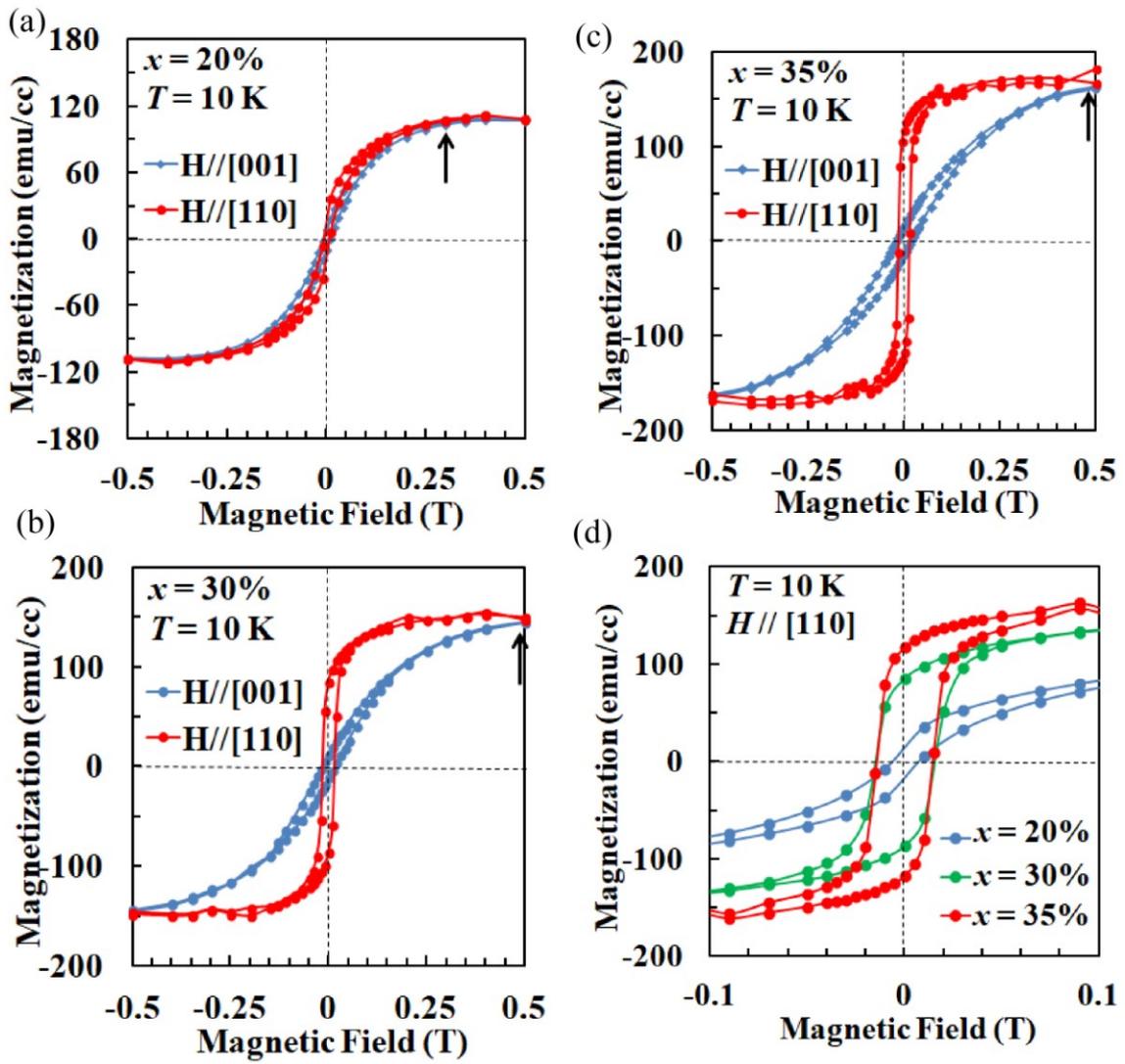

Fig. 2. Tu *et al*

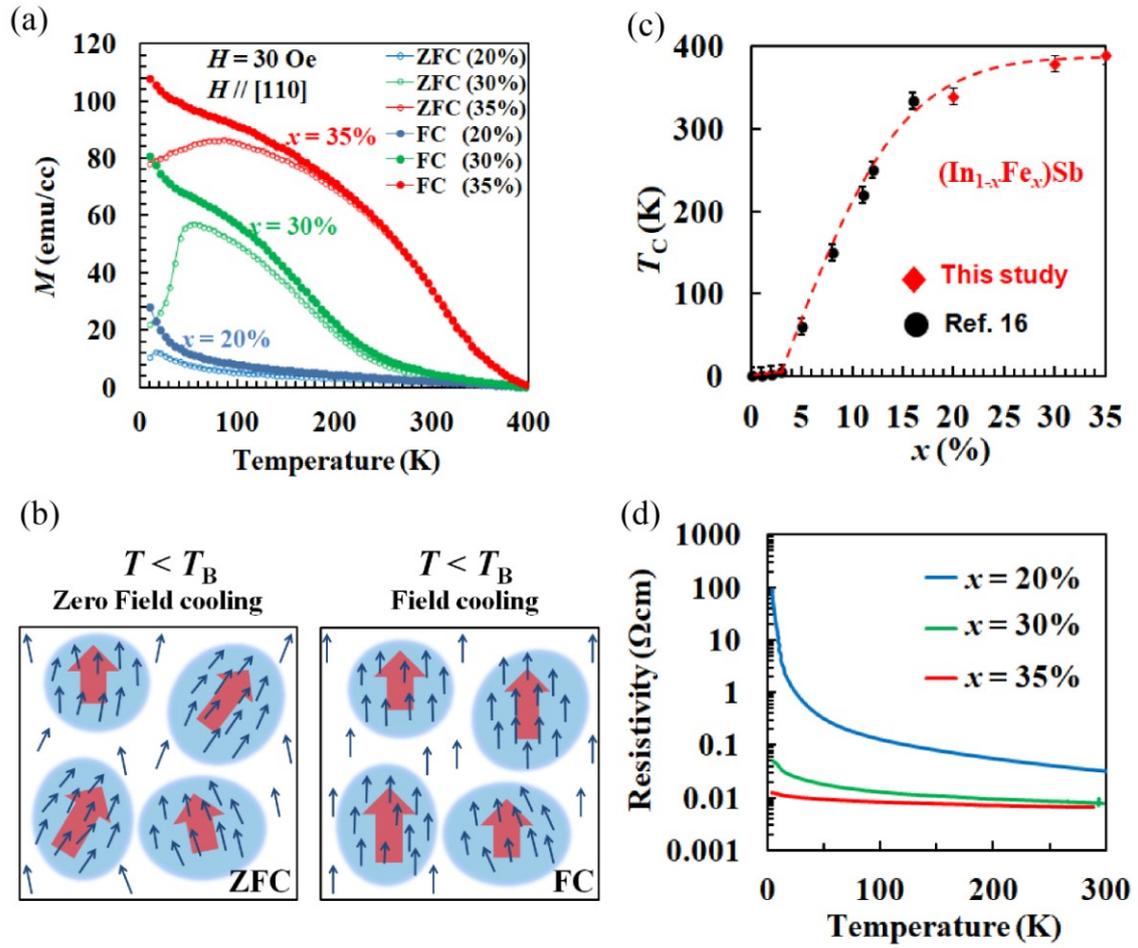

Fig. 3. Tu *et al*

# Supplementary Information

# Heavily Fe-doped *n*-type ferromagnetic semiconductor (In,Fe)Sb with high Curie temperature and large magnetic anisotropy

Nguyen Thanh Tu,[1,2] Pham Nam Hai,[1,3,4] Le Duc Anh,[1,5] and Masaaki Tanaka[1,4]

[1]Department of Electrical Engineering & Information Systems, The University of Tokyo, 7-3-1 Hongo, Bunkyo, Tokyo 113-8656, Japan.

[2]Department of Physics, Ho Chi Minh City University of Pedagogy,
280, An Duong Vuong Street, District 5, Ho Chi Minh City 748242, Vietnam

[3]Department of Electrical and Electronic Engineering, Tokyo Institute of Technology
2-12-1 Ookayama, Meguro, Tokyo 152-0033, Japan.

[4] Center for Spintronics Research Network (CSRN), The University of Tokyo
7-3-1 Hongo, Bunkyo, Tokyo 113-8656, Japan.

[5] Institute of Engineering Innovation, The University of Tokyo
7-3-1 Hongo, Bunkyo, Tokyo 113-8656, Japan.

## 1. Molecular beam epitaxy growth and crystal structure characterization

Our samples were grown on semi-insulating GaAs(001) substrates by low-temperature molecular beam epitaxy (LT-MBE). Figure 1(a) in main text and Fig. S1(a) show the schematic structure of the samples examined in this study. First, we grew a 50-nm-thick GaAs buffer layer and a 10-nm-thick AlAs layer at a substrate temperature ($T_S$) of 550°C to obtain an atomically smooth surface. Next, we grew a 100-nm-thick AlSb buffer layer at $T_S$ = 470°C to reduce the lattice mismatch between (In,Fe)Sb and GaAs. Then, we grew an (In$_{1-x}$,Fe$_x$)Sb layer with $x$ = 20, 30 and 35% at a growth rate of 0.5 µm/h at $T_S$ = 200 - 220°C. The thickness of the (In$_{1-x}$,Fe$_x$)Sb layer was reduced from 30 nm (at $x$ = 20%) to 12 nm (at $x$ = 30 and 35%) to maintain good crystal quality and prevent phase separation. Finally, we grew a 2-nm-thick InSb cap layer to prevent oxidation of the underlying (In,Fe)Sb layer.

During the MBE growth, the crystallinity and surface morphology of the samples were monitored *in situ* by reflection high energy electron diffraction (RHEED). Figures S1(b) – (d) show RHEED patterns taken along the $[\bar{1}10]$ azimuth after the MBE growth of the $(In_{1-x},Fe_x)Sb$ layers of samples B1 – B3 ($x$ = 20, 30 and 35%). The RHEED patterns of (In,Fe)Sb show the zinc-blende type crystal structure without any visible second phase, which are similar to the RHEED patterns of the undoped InSb sample and the $(In_{1-x},Fe_x)Sb$ samples with low $x$ (1 – 16%).[1] This suggests that the (In,Fe)Sb layers ($x$ = 0 – 35%) grown by LT-MBE maintain the zinc-blende type crystal structure.

Figure S2(e) shows X-ray diffraction (XRD) spectra of the $(In_{1-x},Fe_x)Sb$ samples B1 – B3 ($x$ = 20, 30 and 35%). One can see only the (004) diffraction peaks of the GaAs substrate, AlSb buffer and (In,Fe)Sb layers with zinc-blende crystal structures. No other phases such as FeSb and $FeSb_2$ were detected in the XRD spectra. This also suggests that (In,Fe)Sb layers grown by LT-MBE maintain the zinc-blende crystal structure without any visible second phase, which is in good agreement with the RHEED patterns. We take into account the strain effect and estimate the intrinsic lattice constant $a$ of the $(In_{1-x},Fe_x)Sb$ samples with high $x$ from the XRD spectra. The inset of Fig. S1(e) shows the relation $a$ *vs.* $x$. For comparison, $a$ of samples with low $x$ (1 – 16%) was also shown. The lattice constant $a$ of the samples with high $x$ (20 – 35%) does not follow the trend (black dashed line) extrapolated from the relation $a$ *vs.* $x$ of samples with low $x$ (1 – 16%). The reason for this behavior is unclear at this stage. However, a part of Fe atoms may reside at the interstitial sites in heavily Fe-doped $(In_{1-x},Fe_x)Sb$ samples ($x$ = 20 – 35%), which is similar to the case of $(Ga_{1-x},Mn_x)As$ with high Mn concentrations ($x$ > 10%). Therefore, such interstitial Fe atoms may affect the lattice constant of (In,Fe)Sb layers.

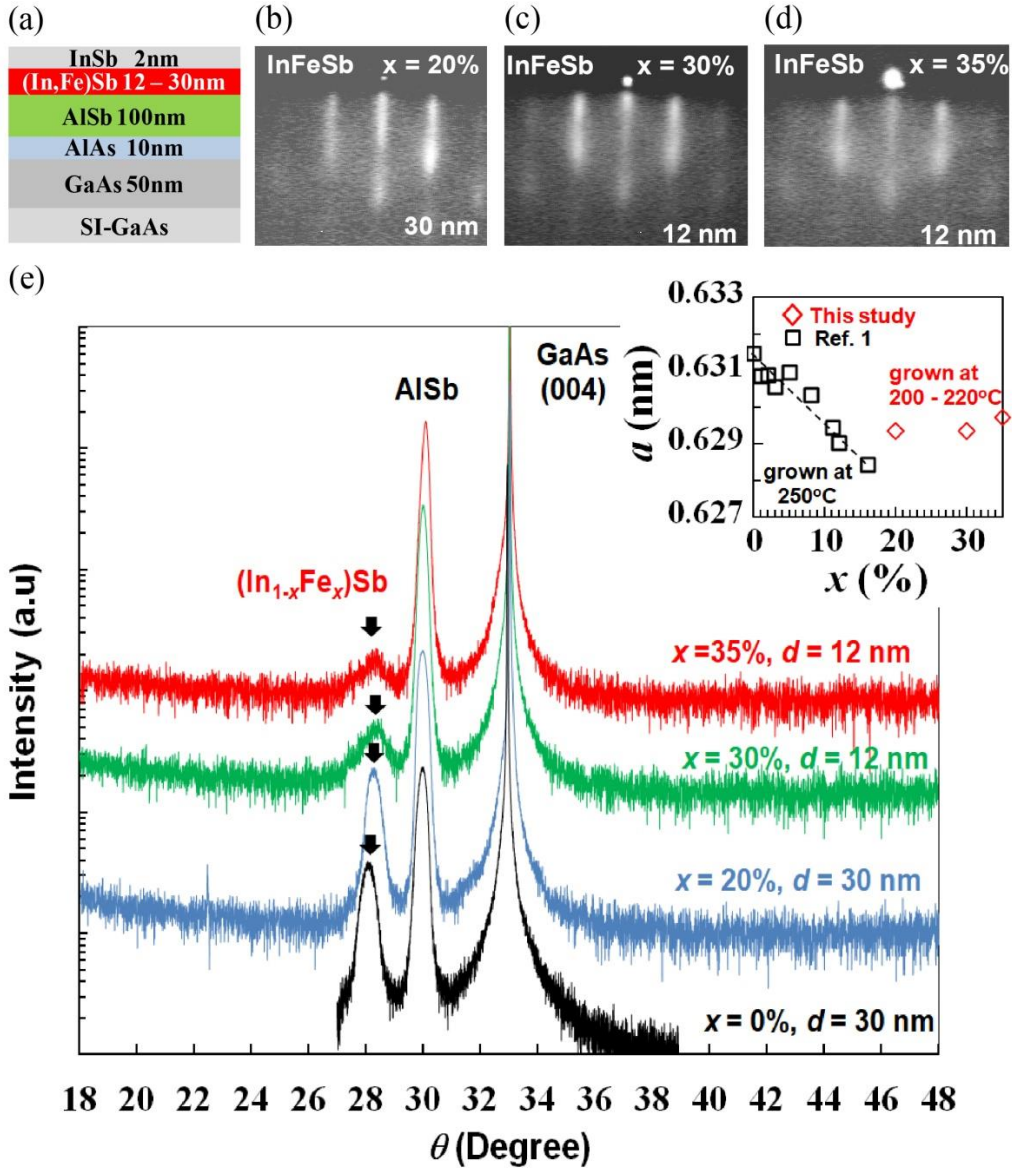

Fig. S1. (a) Schematic sample structure studied in this work. (b) – (d) RHEED patterns taken along the $[\bar{1}10]$ azimuth after the MBE growth of the $(In_{1-x},Fe_x)Sb$ layers for samples B1 – B3 ($x$ = 20, 30 and 35%). (e) X-ray diffraction (XRD) of the $(In_{1-x},Fe_x)Sb$ samples B1 – B3 ($x$ = 20, 30 and 35%, thickness $d$ = 12 – 30 nm) grown on GaAs (001) substrates. An XRD spectrum of the undoped InSb layer ($x$ = 0%) is also shown. The inset of Fig. 1(e) shows Fe concentration ($x$) dependence of the intrinsic lattice constant $a$ of $(In_{1-x},Fe_x)Sb$ measured by XRD.

## 2. Magnetic circular dichroism (MCD) spectroscopy and normalized MCD – $H$ characteristics

In this study, we investigate the magneto-optical properties of the $(In_{1-x},Fe_x)Sb$ samples by magnetic circular dichroism (MCD) spectroscopy in a reflection setup. Here, the reflection MCD intensity is expressed as $(90/\pi)[(R_+ - R_-)/(R_+ + R_-)] \propto (1/R)(dR/dE)\Delta E$. Here, $R_+$ and $R_-$ are the reflectance for $\sigma^+$ and $\sigma^-$ circular polarized light, respectively, $R$ is the total reflectance, $E$ is the photon energy, and $\Delta E$ is the Zeeman splitting energy. Since the MCD intensity is proportional to $dR/dE$ and $\Delta E$, the MCD spectrum directly probes the spin-polarized band structure of the measured material. The MCD spectrum of an intrinsic FMS film would show the spectral features of the host semiconductor with enhanced peaks at their optical critical point energies, whereas the MCD spectrum of a semiconductor film with embedded metallic magnetic precipitates would be broad without any strong peaks related to the host semiconductor band structure. From the MCD spectral shape, we can judge whether the ferromagnetism comes from an intrinsic FMS or from magnetic precipitates.[2,3] Figures 1(c) - 1(e) of the main text show the MCD spectra of the $(In_{1-x},Fe_x)Sb$ samples B1 – B3 ($x$ = 20, 30 and 35%). All the MCD spectra show enhanced peaks at the optical critical point energy $E_1$ and $E_1 + \Delta_1$ of the InSb band structure. We observed no broad background signal in the MCD spectra, which would be observed if there were Fe-related metallic nanoclusters in the InSb matrix. Therefore, our result indicates single phase ferromagnetism in our $(In_{1-x},Fe_x)Sb$ samples.

## 3. Currie-Weiss plot

Figures S3(a) – S3(c) show field cooling (FC, red) magnetization versus temperature ($M$ – $T$) curves of samples B1 – B3 ($x$ = 20, 30 and 35%). Due to the small remanent magnetization, a small magnetic field of 30 Oe was applied in the film plane along the [110] axis during the $M$ – $T$ measurement, which caused a tail up to temperature above $T_C$. Therefore, we estimate $T_C$ of these samples by using the Curie-Weiss plot at high temperatures. In Figs. S3(a) – S3(c), the purple circles show the inverse of the FC magnetization (1/$M$) under a magnetic field of 30 Oe at high temperatures, which follows the Curie-Weiss law. The black broken lines are the fitting with $\chi = \frac{M}{H} = \frac{C}{T-T_C}$, where $\chi$, C, $T$ are the magnetic susceptibility, Curie constant, temperature, respectively. The $T_C$ value is estimated by the intersection of the black broken lines ($\propto \chi^{-1}$) with the horizontal axis. $T_C$ values are listed in the 6th column of Table I and plotted as function of $x$ (red diamonds) as shown in Fig. 3(c) of the main text.

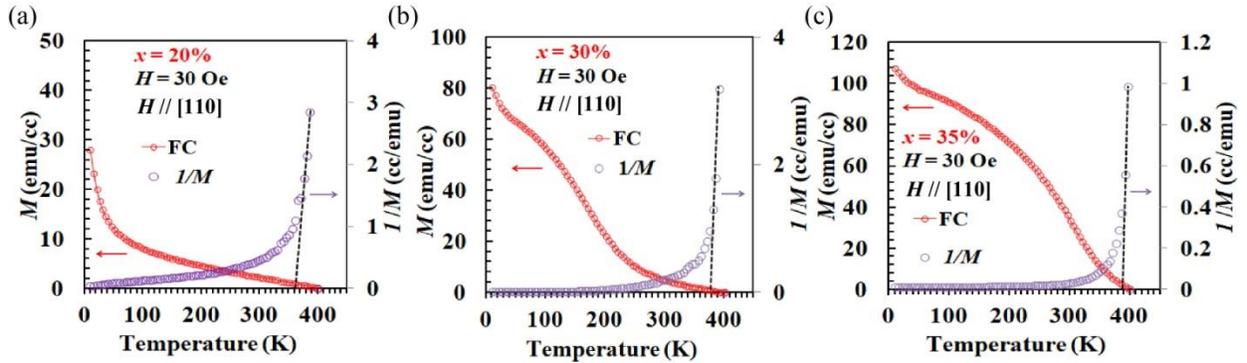

**Figure S3**. (a) – (c) Temperature dependence of the magnetization ($M$ – $T$ curves) of samples B1 – B3 ($x$ = 20, 30 and 35%) measured by SQUID. The samples were cooled from 300 K to 10 K under a magnetic field of 1 T (FC, red dots). After cooling, the magnetization was measured with increasing temperature with a weak magnetic field of 30 Oe applied perpendicular to the plane along the [001] axis. The purple circles show the inverse of the FC magnetization (1/$M$) under a magnetic field of 30 Oe at high temperatures, which follows the Curie-Weiss law.

## 4. Hall resistance *vs*. perpendicular magnetic field ($R_{Hall} - H$) characteristics

Figures S4(a) –4(c) show the Hall resistance *vs*. perpendicular magnetic field ($R_{Hall} - H$) curves of the samples B1 – B3 ($x$ = 20, 30 and 35%) at various temperatures 50 – 320 K. To eliminate the magnetoresistance contributions that are even functions of $H$, the odd-function contributions are extracted from the raw Hall data and plotted in Figs. S4(a) – (c). We cannot estimate the hole concentration due to the influence of the anomalous Hall effect even at 320 K. $R_{Hall}$ are dominated by the anomalous Hall effect (AHE) with open hysteresis even at 320 K, which confirms again ferromagnetic order above room temperature. The high field saturation of $R_{Hall}$ indicates again that the easy axis is in the film plane. The $R_{Hall} - H$ characteristics agree well with the MCD – $H$ and $M - H$ characteristics (see Supplementary Information Section 5 for detailed comparison between the characteristics measured by AHE, SQUID and MCD).

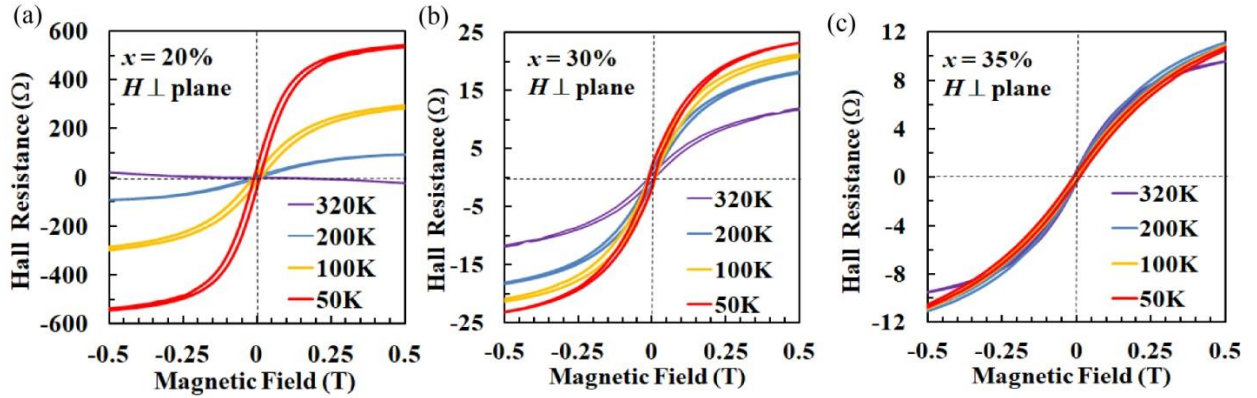

Fig. S4. (a) – (c) Hall resistance *vs*. magnetic field $H$ applied perpendicular to the film plane of the (In,Fe)Sb films of samples B1 – B3 ($x$ = 20, 30 and 35%), respectively, measured at various temperatures 50 – 320 K. To eliminate the magnetoresistance contributions that are even functions of $H$, the odd-function contributions are extracted from the raw Hall data and plotted in (a) – (c).

## 5. Comparison between SQUID, $R_{Hall}$, and MCD measurements

Figures S5(a) – (c) show normalized hysteresis curves (magnetic field dependences) of samples B1 – B3 ($x$ = 20, 30 and 35%), respectively, measured by MCD and SQUID at 10 K. Here, we cannot measure AHE at 10 K due to its high resistance. Figures S5(d) – (f) show normalized hysteresis curves of samples B1 – B3 ($x$ = 20, 30 and 35%), respectively, measured by $R_{Hall}$ (dominated by AHE) and MCD at 50 K. One can see that hysteresis curves measured by different methods perfectly agree with each other, indicating the single ferromagnetic phase in these samples, which come from the zinc-blende ferromagnetic semiconductor phase, not from second-phase precipitations if any.

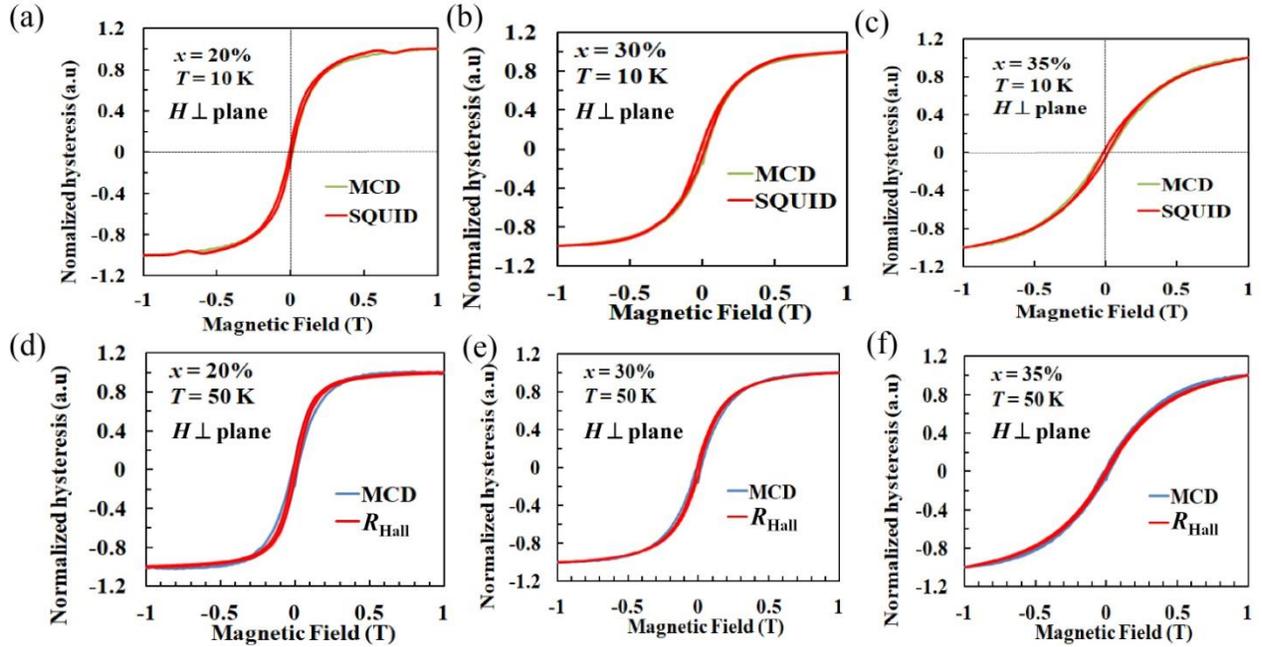

Figure S5 (a) – (c) show normalized hysteresis curves of samples B1 – B3 ($x$ = 20, 30 and 35%), respectively, measured by MCD and SQUID at 10 K. Here, we cannot measure $R_{Hall}$ at 10 K due to its high resistance. (d) – (f) show normalized hysteresis curves of samples B1 – B3 ($x$ = 20, 30 and 35%), respectively, measured by $R_{Hall}$ (dominated by AHE) and MCD at 50 K